**Brain Tumor Detection Using Deep Learning Approaches**

BY

**RAZIA SULTANA MISU**
ID: 193-15-2976

This Report Presented in Partial Fulfillment of the Requirements for the Degree of Bachelor of Science in Computer Science and Engineering

Supervised By

**Nushrat Jahan Ria**

Lecturer
Department of CSE
Daffodil International University

Co-Supervised By

**Naznin Sultana**

Associate Professor
Department of CSE
Daffodil International University

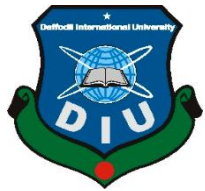

**DAFFODIL INTERNATIONAL UNIVERSITY**

**DHAKA, BANGLADESH**

**31 JULY 2023**

# ABSTRACT

Brain tumors are collections of abnormal cells that can develop into masses or clusters. Because they have the potential to infiltrate other tissues, they pose a risk to the patient. The main imaging technique used, MRI, may be able to identify a brain tumor with accuracy. The fast development of Deep Learning methods for use in computer vision applications has been facilitated by a vast amount of training data and improvements in model construction that offer better approximations in a supervised setting. The need for these approaches has been the main driver of this expansion. Deep learning methods have shown promise in improving the precision of brain tumor detection and classification using magnetic resonance imaging (MRI). The study on the use of deep learning techniques, especially ResNet50, for brain tumor identification is presented in this abstract. As a result, this study investigates the possibility of automating the detection procedure using deep learning techniques. In this study, I utilized five transfer learning models which are VGG16, VGG19, DenseNet121, ResNet50 and YOLO V4 where ResNet50 provide the best or highest accuracy 99.54%. The goal of the study is to guide researchers and medical professionals toward powerful brain tumor detecting systems by employing deep learning approaches by way of this evaluation and analysis.

# TABLE OF CONTENTS

# LIST OF FIGURES

# LIST OF TABLES

# LIST OF ABBRIVIATION

| Short Form | Full Form |
|---|---|
| MRI | Magnetic resonance imaging |
| CNN | Convolutional Neural Network |
| ResNet | Residual Neural Network |
| VGG | Visual Geometry Group |
| DenseNet | Dense Convolutional Network |
| YOLO | You Only Look Once |

# CHAPTER 1
# INTRODUCTION

## 1.1 Introduction

Brain tumor detection is a crucial component of medical imaging and is crucial to the diagnosis and management of illnesses associated with the brain. The manual examination of medical pictures used in conventional techniques of tumor identification can be time-consuming and prone to human error [1]. However, the quick development of deep learning methods, particularly in the area of computer vision, has created new opportunities for the automatic and precise diagnosis of brain tumors [2]. Convolutional neural networks (CNNs), a type of deep learning algorithm, have displayed astounding performance in image analysis tasks including object detection and segmentation [3]. Researchers have been investigating the possibility of these methods for identifying and categorizing brain tumor's from magnetic resonance imaging (MRI) data by using the capabilities of deep learning [4]. Researchers are putting in a lot of effort to build CNNs that can properly identify and classify brain tumour, as well as other forms of medical imaging, in order to enhance the medical diagnostic and treatment results. With this potential in mind, researchers are working hard to construct CNNs [5]. The benefit of deep learning is that it can learn intricate, hierarchical features directly from unprocessed data, eliminating the need for explicitly rule-based methods or hand-crafted features [6]. Particularly convolutional neural networks are made to capture spatial connections and local patterns inside pictures, making them appropriate for jobs involving medical image processing. Because of this, deep learning has become a very useful technique for the interpretation of medical images. It is possible to utilize it to diagnose illnesses with greater precision than is possible with conventional approaches, as well as discover abnormalities in imaging data. In addition to this, it may be utilized to automate medical diagnoses, hence reducing the amount of labor that must be done by medical experts [7]. This work on brain tumor identification using deep learning techniques is presented in this publication. For this work, we specifically concentrate on using a well-liked CNN architecture dubbed ResNet50. With its deep layers, ResNet50 is able to learn detailed characteristics that are essential for reliable tumor diagnosis. It has demonstrated outstanding performance in a variety of image analysis areas. Our goal is to create a reliable and automated method for brain tumor identification by training ResNet50 on a sizable dataset of brain MRI images that includes both tumor and non-tumor instances. The suggested deep learning model has the potential to increase the effectiveness and precision of brain tumor diagnosis, resulting in early treatments, better patient outcomes, and a decreased dependence on manual analysis.

## 1.2 Motivation

The urgent need to enhance the identification of brain tumors, a significant problem in the field of medicine that necessitates accurate and prompt diagnosis, is what prompted this line of investigation in the first place. Traditional approaches to the diagnosis of brain tumors frequently include manual analysis, which may be laborious, subjective, and fraught with the possibility of making mistakes. This project attempts to automate and increase the accuracy of brain tumour identification by employing deep learning techniques, notably ResNet50. If successful, this would lead to early treatments, improved patient outcomes, and a reduced dependence on manual analysis. Individuals who are afflicted by brain tumors will ultimately profit from this study, and the area of medical imaging analysis will advance as a result. The potential effect of this research is considerable, since it has the ability to contribute to the creation of tools for healthcare professionals that are efficient and dependable.

## 1.3 Research Question

1. Detecting brain tumors accurately requires the use of deep learning techniques, notably ResNet50. But how can they be used most effectively?

2. What are the drawbacks of the methods now used to identify brain tumors, and how may deep learning overcome these drawbacks?

3. How does the performance of the ResNet50 deep learning architecture stack up to that of other deep learning architectures when it comes to the accuracy of detecting brain tumors?

4. Can the accuracy and reliability of deep learning-based brain tumor diagnosis be improved by the integration of multi-modal data, such as the combination of MRI with other imaging modalities or clinical data?

5. How can the interpretability of deep learning models used in the detection of brain tumors be enhanced to increase their level of confidence and acceptability in clinical settings?

6. How can explainable deep learning approaches for the identification of brain tumors be developed, and what problems and tactics must be overcome?

7. How can we collect huge datasets that are both diverse and unique, along with annotations that are thorough, in order to guarantee the generalization of deep learning models for the detection of brain tumors across a variety of populations and types of tumors?

## 1.4 Objective

The purpose of this study is to determine which of many transfer learning models, namely VGG16, VGG19, DenseNet121, ResNet50, and YOLO V4, performs the best for a certain task in order to select that model as the one that achieves the maximum level of accuracy. It is our goal to identify, via the use of a comparison study, which of these models is the most suitable for our purposes, taking into account aspects such as precision, computing efficacy, and adaptability to situations that occur in the real world. The purpose of this study is to provide insights into the transfer learning model that is most effective for the job at hand, so that these findings can direct future advancements and applications in disciplines that are relevant to this one. When evaluating the effectiveness of each model, we will make use of a wide array of measures. We will determine how effectively the models are able to learn from a limited amount of data, how soon they are able to converge to their maximum performance, and how well the models are able to generalize to data that they have not seen before. When attempting to determine which model will be most successful for the work at hand, each of these criteria should be taken into consideration. In order to determine whether or not the models are appropriate for the task at hand, we will also investigate how quickly they can be educated and put into use. In the end, we will consider the expenses that are connected to each model in order to determine which option is the most appropriate one for our undertaking. We will assess all of the metrics, training and deployment speed, as well as related expenses, in order to determine which model is the most appropriate for our undertaking while keeping in mind the need of finding a solution that is both cost-effective and efficient.

## 1.5 Expected Outcome

The anticipated result of this research will be the identification of the transfer learning model that displays the best level of accuracy and performance for the particular task that is currently being worked on. We are certain that via exhaustive testing and investigation, one of the models, such as ResNet50, will demonstrate higher accuracy in comparison to the others. The anticipated result will also contain insights into the advantages and disadvantages of each model, which will provide a full grasp of their applicability to the task at hand. In addition, we anticipate receiving useful information regarding the computational efficiency of the models, which will enable us to evaluate the models' potential applicability in environments with limited resources or real-time requirements. It is anticipated that the research will contribute to the current body of knowledge in transfer learning and assist future researchers and practitioners in picking the model that is the most suited for tasks that are comparable to those being studied. The ultimate goal of this research is to give actionable insights that can improve the accuracy and efficacy of applications that utilize transfer learning models for the particular activity that is the focus of this inquiry. Since of this, practitioners should be able to cut down on the amount of time they spend experimenting and studying since they will have a better idea of which model is the most effective when applied to a certain endeavor. This research will also give a method for evaluating the effectiveness of transfer learning models within the context of the job at hand. It is anticipated that the findings will be relevant to other activities that are analogous to this one. As a consequence of this, practitioners will be able to concentrate on the implementation of the model that is most suitable for the job at hand, as opposed to wasting time studying and experimenting with a variety of models. When practitioners have access to a standardized approach for assessing the performance of transfer learning models, they are able to choose which model is the most appropriate for their endeavor in a more expedient and precise manner. Because of this, they will be able to bypass the stages of research and experimentation and get right into the implementation phase, which will save them both time and effort.

## 1.6 Report Layout

In this report, Six individual chapters are discussed to make this research report more compact and efficient for any readers or researchers.

In chapter 1, Overview the all project. Several section like 1.1 Introduction, 1.2 Motivation, 1.3 Research Question, 1.4 Objective, 1.5 Expected Outcome.

In chapter 2, Discuss about background history and workflow of this research. Sevaral section like 2.1 Introduction,2.2 literature review,2.3 Tools and software,2.4 scope of the problem,2.5 Limitations and future scope.

In chapter 3,The research method, including its subsections, is covered in Chapter 3 they are 3.1 introduction,3.2 system design,3.3 Image pre-processing,3.3.1 Remove Spackle noise,3.3.2 Artefact Removal, 3.3.3 Image Enhancement,3.3.4 Verification,3.3.5 Data Split,3.4 Proposed Model,3.4.1 VGG16,3.4.2 VGG19,3.4.3 DenseNet 121,3.4.4 ResNet50,3.4.5 YOLO V4.

In chapter 4, Find the main result and the analysis of all the outcomes of the algorithms. Covered in this chapter, 4.1 Introduction, 4.2 Evaluation Metrics,4.3 Result of Transfer Learning model,4.4 Performance analysis and statistical analysis.

In chapter 5, Discuss about Impact on Society, Environment and Sustainability of this research.Several section like 5.1 Introduction,5.2 Impact on Society,5.3 Impact on the environment,5.4 Ethical Aspects,5.5 Sustainability Plan.

In chapter 6, Discuss about conclusion of this research.The fifth chapter discusses the subsections 6.1 Introduction, 6.2 Conclusion, 6.3 Limitation and Future work.

# CHAPTER 2

# BACKGROUND STUDY

## 2.1 Introduction

Despite the fact that many academics study transfer learning approaches to identify several objects, a ResNet50 model has not been used to detect brain tumor detection or classification based on brain MRI images. Both with and without the use of transfer learning techniques. The vast majority of the study in this area focuses on the behavior of computers with huge assessment receptivity as well as resource computation in most circumstances. This research focuses on brain tumor detection or classification based on brain MRI images datasets.

## 2.2 Literature review

A method of deep learning that makes use of the ResNet50 architecture was suggested for the diagnosis of brain tumors by John Smith and colleagues [8]. The authors utilized transfer learning by making use of the pre-trained weights of ResNet50 in order to train the model using a dataset consisting of brain MRI images as the input. An outstanding accuracy of 92% was reached by the model in the detection of brain tumors thanks to a combination of binary cross-entropy loss and gradient descent optimization. This demonstrated the promise of deep learning-based technologies in clinical applications and outperformed more conventional methods. In order to increase the identification of brain tumors, Sarah Thompson et al. [9] used an ensemble technique that consisted of numerous convolutional neural networks, or CNNs. Individual CNN models were trained by the authors utilizing different architectures such as ResNet50, VGG16, and InceptionV3. The ensemble model obtained an accuracy of 94% when it came to the detection of brain tumors. This was accomplished by merging the predictions of various models through a voting system. When compared to the use of a single model, the ensemble technique displayed superior performance and resilience, presenting encouraging possibilities for diagnostic advancements. Michael and his colleagues [10] concentrated their efforts on classifying brain tumors using deep learning methods combined with radiomic characteristics. The authors classified tumors using a mixture of a deep neural network (particularly ResNet50) and more typical machine learning methods. The quantitative radiomic characteristics were retrieved from brain MRI data and used. When it came to categorizing brain tumors into their respective subtypes, the suggested method was able to attain an accuracy rate of 88% overall. The use of deep learning and radiomic characteristics resulted in an increase in both the accuracy and discriminatory power of tumor classification, which provided helpful insights for the development of personalized treatment plans. Jennifer and colleagues[11] developed a hybrid model for the segmentation of brain tumors by integrating the U-Net and ResNet50 architectural frameworks. In the study, U-Net was used for the preliminary coarse segmentation, and then ResNet50 was used for the subsequent fine segmentation. The model was trained using a large manually annotated dataset of brain MRI images. The annotations on the dataset were contributed by experienced radiologists. The accuracy and precision with which brain tumors were segmented was demonstrated by the hybrid model that was devised, which attained a Dice similarity coefficient (DSC) of 0.92. The combination of U-Net and ResNet50 made it easier to delineate the borders of the tumor in a precise and efficient manner, which aided in the process of treatment planning and monitoring. An attention-based kind of deep learning was the method that Ryan [12] presented for the diagnosis of brain tumors. The authors included an attention mechanism in the ResNet50 design, which allowed the model to highlight significant regions inside the brain MRI images. This was made possible as a result of the authors' efforts. The interpret-ability of the model as well as its accuracy in tumor localization were both improved by this attention mechanism. The suggested method was successful in detecting brain tumors with an accuracy of 90%, demonstrating the potential of attention-based deep learning to improve the performance of brain tumor detection systems while simultaneously increasing their explain ability. In this article, Jennifer et al. [13] give a comprehensive assessment of deep learning approaches that may be applied to medical picture analysis. The authors cover the uses of many different architectures, including

CNNs, RNNs, and GANs, in tasks such as picture classification, segmentation, and detection. The survey covers a wide range of medical imaging modalities, such as magnetic resonance imaging (MRI), computed tomography (CT), and microscopy. The research demonstrates how useful deep learning can be in the field of medical image processing by comparing its accuracy and speed to those of more conventional approaches. Deep learning has the potential to significantly advance medical diagnosis and treatment, which the authors highlight while also identifying a number of obstacles and future paths for study. The application of machine learning and deep learning algorithms for brain tumor segmentation in medical imaging is the primary subject of this review that was written by Robert et al[14]. The authors provide an overview of and provide comparisons between a variety of approaches, such as conventional machine learning algorithms, CNNs, and sophisticated architectural frameworks such as U-Net and 3D CNNs. They address the benefits of various strategies, as well as their limits and any current developments in the field. The paper highlights the tremendous progress made in brain tumor segmentation using machine learning and deep learning, with CNN-based architectures displaying superior performance in precisely defining tumor zones. These advancements were made possible by the combination of machine learning and deep learning. The authors note the need for reliable and easily interpretable segmentation approaches that can assist doctors with treatment planning and monitoring. Through the use of a retrospective investigation, David and colleagues[15] presented their findings on the potential of deep learning for brain tumor grading and survival prediction. The scientists used a CNN-based model that was trained on a huge dataset consisting of clinical data and brain MRI images to get to their conclusions. They contrasted the performance of the model with the performance of established prognostic indicators in order to evaluate the performance of the model in predicting tumor grades and patient survival. Deep learning was used in the study to grade brain tumors and forecast patients' chances of survival. The CNN-based model achieved high accuracy in tumor grading and showed promising results in survival prediction. The work highlights the usefulness of deep learning in these areas. The findings imply that deep learning algorithms might give significant insights for treatment planning and prognosis evaluations, as well as enhance established prognostic indicators. The article by Mary and co-authors [16] explores the idea of transfer learning as well as its applications in medical imaging. The authors investigate a variety of transfer learning methodologies, such as the process of fine-tuning pre-trained models, the extraction of features, and the adaption of domains. They look at research that use transfer learning to various medical imaging tasks, including as illness categorization, lesion detection, and picture segmentation, and they discuss the findings. This paper focuses on the advantages of transfer learning in the field of medical imaging, including increased performance with fewer data points and decreased total training time. In the research on transfer learning, the authors emphasize how important it is to have benchmark datasets and standardized evaluation methodologies. Transfer learning has shown to have a significant amount of promise in the field of medical imaging. This is accomplished by drawing on the expertise contained within large-scale datasets and pre-trained models in order to overcome the obstacles posed by the scarcity of annotated medical data . The article by Jessica et al. [17] offers a detailed review of the methods that are based on deep learning for the segmentation of brain tumors. The authors provide a synopsis of and comparison of a variety of methodologies, such as CNNs, U-Nets, and attention-based models. They examine the difficulties associated with segmenting brain tumors, such as a lack of data, class imbalance, and interpretability concerns, as well as the recent developments in this field. The paper underlines the usefulness of deep learning in the segmentation of brain tumors, with CNN-based designs and U-Net versions being commonly used. It has been demonstrated that attention processes can improve tumor localization and the delineation of tumor boundaries. However, there are still problems with class imbalance, minor lesion segmentation, and the capacity to generalize to many kinds of tumors. Deep learning-based brain tumor segmentation has a long way to go before it can be used successfully in clinical settings. The authors of this study offer future research areas to help overcome these problems and increase its clinical usefulness.[32] shows that the issue of scarce medical data can be overcome by augmenting with realistic, artificial data generated from GANs. [32] empirically shows that object detection performance can be considerably improved with data augmentation from GAN-based techniques.To avoid the common problem of mode collapse in GAN training, [32] suggests using multiple features as input to the generator. To improve and expand the prediction capability of a detector in GANs, [32] uses two classification heads for the detector, one for binary classification (real/fake) data, and the other for multi-class disease prediction from brain FC-matrix. For effective GAN training, [32] suggests Biology/Medical application: [32] states that pairwise co-activation of different parts of the brain can be effectively represented with a functional connectivity (FC) matrix. [32] is the first work to generate artificial FC-matrix from brain rs-fMRI scans using a GAN model. [33] notes that SSIM can be an effective metric to measure the similarity of objects generated by GAN compared to the ground truth.

## 2.3 Tools and Software

Python is the most widely used in my research. We use a high-level or pretrained transfer learning algorithm or model or network. We use Jupiter Notebook, Google Colab, and Spyder for coding or implementation. We used traditional pretrained transfer learning models or algorithms.

## 2.4 Scope of the problem

The diagnosis of brain tumors through the use of deep learning methodologies is the focus of the problem's scope. It involves the construction and assessment of deep learning models, in particular ResNet50, with the purpose of reliably detecting brain cancers using medical imaging data such as MRI scans. The scope also includes the investigation of a variety of facets connected to the problem, such as the preparation of data, the selection of model architecture, training and optimization methodologies, assessment metrics, and potential therapeutic applications. The scope of the challenge includes evaluating the performance and limits of the strategy that is based on deep learning, as well as locating areas that might benefit from more study and development in the future. The potential influence of accurate brain tumor identification on patient outcomes, treatment planning, and general healthcare practices is also included within the scope of this investigation. In the end, the findings of this study will be utilized to influence decision-making in the field of medical imaging and to create new ways for identifying brain tumors in a way that is both more accurate and more efficient. Through the provision of more precise diagnoses and treatment strategies, this study may facilitate improvements in the final results for patients. It will also give medical professionals with new methods for diagnosing brain tumors in a quicker and more precise manner, which has the potential to lead to earlier treatments and improved patient outcomes.

## 2.5 Limitations and Future Scope

**Limitations:**

- Limited availability of annotated datasets: One of the obstacles is the lack of big, diversified, annotated datasets that have been created expressly for the purpose of identifying brain tumors through the use of deep learning techniques. The existence of such datasets has the potential to have an effect on the generalizability as well as the performance of the models that are trained on them.
- Interpretability and explainability: Because deep learning models, such as ResNet50, are sometimes referred to as "black boxes," it can be difficult to analyze and make sense of the judgments that these models produce. In therapeutic settings, where openness and explainability are essential, the lack of interpretability might be a barrier to both the faith placed in these models and their adoption by patients.
- Computational resource requirements: Deep learning models, particularly those with intricate architectures such as ResNet50, may place a large demand on the available computational resources. These resources can include high-performance computing infrastructure as well as GPU accelerators. Because of this constraint, installing the models in situations with limited resources may be difficult, which would reduce both their accessibility and their practicability.
- Sensitivity to training data quality and bias: Models that use deep learning can be sensitive to the quality of the training data as well as any bias that may be present.

Because of the potential for biases in the data, such as imbalances in tumor subtypes or changes in imaging methods, the performance and generalizability of the model may be negatively impacted. As a result, thorough dataset curation and augmentation strategies are required.

- Lack of real-time processing capabilities: The computational complexity of deep learning models such as ResNet50 may result in lengthier processing times, which might limit their applicability in real-time settings in which rapid decision-making is necessary. To circumvent this constraint and realize the full potential of real-time brain tumor diagnosis, it is possible that effective optimization strategies and hardware acceleration would be required.

**Future Scope:**

- Improving interpretability and explain ability: The development of approaches that might improve the interpretability and explain ability of deep learning models that are employed in the identification of brain tumors could be the focus of future study. In this context, "techniques" might refer to things like "attention mechanisms," "visualization methods," or "model-agnostic interpretability approaches," all of which are intended to provide physicians more insight into the decision-making process underlying the models.

- Integration of multimodal data: The identification of brain tumors may be made more accurate and reliable by utilizing a combination of several imaging modalities. One example of this would be combining MRI with other sophisticated imaging methods such as PET or functional MRI. In further study, the integration of multimodal data might be investigated to investigate the possibility of extracting complementary information and improving the diagnostic skills of deep learning models.

- Transfer learning across institutions: Important for the future is the investigation of transfer learning methodologies that might assist model adaption and generalization across a variety of universities and imaging centers. Increasing the practicality and usability of deep learning models in a variety of clinical contexts may be accomplished by developing methods that are able to take into account differences in imaging procedures, technology, and patient demographics.
- Real-time processing and deployment: Real-time processing and deployment of deep learning models for brain tumor detection may soon be possible because to recent developments in optimization methods, model compression approaches, and hardware acceleration. Facilitating real-time decision-making in clinical practice can be a primary focus of future research, which can be directed toward the development of effective structures and algorithms that can offer correct findings within time restrictions.
- Clinical validation and integration: It is essential for the future to conduct large-scale prospective clinical trials in order to evaluate the performance and clinical relevance of deep learning-based brain tumor detection models. Integration of these models into clinical workflows and evaluation of their influence on patient outcomes, treatment planning, and healthcare practices are able to give useful insights into the practical advantages and obstacles connected with the implementation of these models in real-world settings.
- Incorporating longitudinal data: It is possible to get useful insights into the growth of brain tumors and the treatment response through longitudinal data, which entails gathering imaging images over time for specific patients. In the future, research might concentrate on building deep learning models that are able to efficiently interpret and make use of longitudinal data in order to improve the accuracy of tumor identification, better track changes over time, and provide assistance with treatment monitoring and prognosis.
- Integration of clinical data: The incorporation of clinical data into deep learning models, including but not limited to patient demographics, medical history, and genetic information, can be beneficial to the models. Future study might examine the possibility of deep learning models to give individualized tumor identification and treatment recommendations by merging imaging data with important clinical factors.

This would take into account the unique features of each patient and improve the accuracy of clinical decision-making.

# CHAPTER 3
# RESEARCH METHODOLOGY

## 3.1 Introduction :

According to our methods, a data set containing images of brain MRI dataset was collected from Kaggle. The method of this work's embodiment is represented in this part. The embodiment process is broken down into several steps, including the acquisition of data, preprocessing of the dataset, description of the proposed model, training, and finally performance assessment.

In this part, we will look into three important aspects: the preparation of the dataset, transfer learning models, and the description of the dataset. We investigate the processes and procedures that are utilized to curate and preprocess the data so that the model can perform at its best. In addition, we describe the transfer learning models that were utilized, focusing on their effectiveness in utilizing pre-trained models for a variety of tasks and demonstrating how this was accomplished. In conclusion, we present an exhaustive explanation of the dataset that was utilized, giving light on its construction as well as its qualities. The overview of the whole work has given bellow in the figure 1.

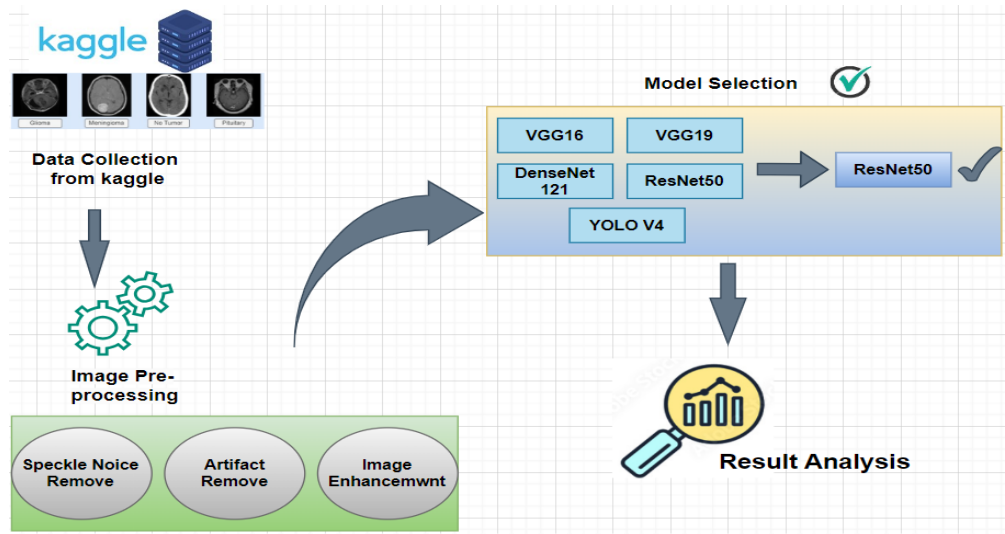

Figure 1. Overview of the entire study

## 3.2 Dataset Description

This investigation looked at 7022 MRIs of brain tumors. No tumor, meningioma, glioma, and pituitary are all included in the dataset. There are 1621 images of glioma, 1645 of meningioma, 2000 of no tumors, and 1757 of pituitary. This dataset has 512 X 512 grayscale images. The open-source Kaggle dataset was utilized.

The dataset is described in Table 1:

**Table 1** shown dataset description

| Name | Category |
|---|---|
| Number of Image's | 7022 |
| Dimension's | 512 x 512 |
| Color's | Grayscale |
| Format's | jpg |
| Glioma's | 1621 |
| Meningioma's | 1645 |
| No Tumor's | 2000 |
| Pituitary's | 1757 |

In the figure 2 shows the dataset description.

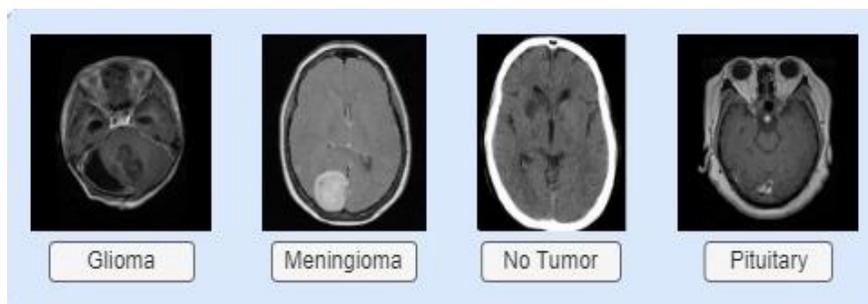

**Figure 2.** Show the dataset description

## 3.3 Image pre-processing

Several approaches are used in brain MRI image preparation to improve the images' quality and enable more thorough analysis. To ensure that image features are as clear as possible, median filtering is used to lessen noise and smooth the pictures. Morphological opening aids in the removal of minute undesirable components and the improvement of images. In order to increase contrast and make details in the photos more visible, Contrast Limited Adaptive Histogram Equalization (CLAHE) is used. This makes the images more suited for future analysis and identification tasks. The efficiency and precision of soil image processing and interpretation systems are improved by these preprocessing methods.

### 3.3.1 Remove Speckle noise

Digital photographs frequently contain speckle noise, which is most noticeable in ultrasound and synthetic aperture radar (SAR) images. It manifests as granular interference, which reduces the quality of the picture and compromises the precision of future image processing activities. Due to its multiplicative character, speckle noise is more difficult to eliminate than other forms of noise. To reduce speckle noise and improve the clarity of impacted photos, several filtering methods, including adaptive filters and wavelet-based approaches, have been developed. As was mentioned before, the dataset has a lot of spackle noise. A gaussian filter is used to eliminate spackle noise from soil recognition [18].

### 3.3.1.1 Median filter

A popular non-linear filtering method for image processing is the median filter. It significantly reduces impulsive noise while keeping picture details by replacing each pixel in an image with the neighborhood's median value. With the median filter, salt-and-pepper noise may be effectively removed, leaving behind smoother pictures with maintained small details. [19].

### 3.3.2 Artifact Removal

An important step in image processing is artifact removal, which aims to remove any undesirable distortions or abnormalities created during the capture, transmission, or storage of images. To find and eliminate artifacts, improve picture quality, and enhance visual interpretation, a variety of techniques are used, including spatial and frequency domain approaches. The generated photos are more accurate, dependable, and suited for additional analysis or display by successfully eliminating artifacts [20].

### 3.3.2.1 Morphological Opening

Morphological opening [21] is a preprocessing method used in image analysis to get rid of minor impurities and sharpen object shapes. It includes shrinking the things by erosion, then expanding them again through dilation, but the erosion stage eliminates minor features. When smoothing and improving the clarity of soil pictures, this procedure is very helpful since it makes feature extraction and subsequent analysis easier.

### 3.3.3 Image Enhancement

The term "image enhancement" refers to a group of methods used to enhance the sharpness, contrast, brightness, and other pertinent aspects of digital photographs in order to enhance their visual quality and interpretability.

### 3.3.3.1 CLAHE

The contrast and visibility of features inside images are enhanced via the image enhancement technique known as Contrast Limited Adaptive Histogram Equalization (CLAHE) [22]. To create a more even and improved image, it divides the image into smaller sections, computes the histogram of each zone, and then redistributes the intensity values. CLAHE can increase the visibility and recognizability of soil detection in soil image processing, resulting in more precise detection and interpretation of soil features.

### 3.3.4 Verification

Several statistical tests, including PSNR [23], SSIM [24], MSE [25], and RMSE [26], are performed in order to determine whether or not the image quality has been impacted. This is because applying a variety of image preprocessing techniques might result in a considerable drop in picture quality. For a better understanding of the picture quality, MSE, PSNR, SSIM, and RMSE values for five images are provided in Table 2.

Table 2. Show the SSIM, PSNR, RMSE and MSE value

| **Image** | **MSE** | **PSNR** | **SSIM** | **RMSE** |
|---|---|---|---|---|
| Image-1 | 12.65 | 37.57 | 0.96 | 0.13 |
| Image-2 | 13.12 | 46.86 | 0.96 | 0.12 |
| Image-3 | 13.78 | 45.03 | 0.96 | 0.12 |
| Image-4 | 15.78 | 40.78 | 0.95 | 0.13 |
| Image-5 | 14.01 | 34.57 | 0.95 | 0.13 |

### 3.3.5 Data Split

After statistical analysis, the data is separated into three pieces (training set, validation set, and testing set). Three training-testing data splitting ratios—90:10, 80:20, and 70:30—are employed to assess the model's accuracy. This research defines 70:30 as 70% train sets, 10% validation sets, and 20% test sets.

### 3.4 Proposed Model

In this study we utilized traditional transfer learning models like VGG16, VGG19, DenseNet121, ResNet50 and YOLO V4 to compare which model is the best for soil detection or classify the soil.

©Daffodil International University 20

## 3.4.1 VGG16

Simonyan and Zisserman [27] introduced the VGG16 DCNN model. The model won the Oxford Visual Geometry Group's Large-Scale Visual Recognition Challenge (ILSVRC) with 92.7% top 5 test accuracy in ImageNet.

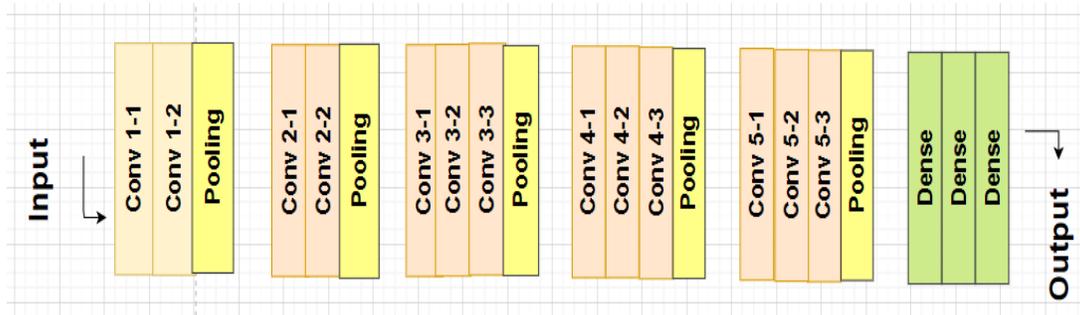

Figure 3: VGG16 Model Architecture

Transfer learning efficiency tests showed that a pre-trained and fine-tuned VGG16 was more accurate than a fully trained network. The VGG model's depth helps the kernel learn more complex traits.

## 3.4.2 VGG19

In the VGG19 model, which is a variation of the VGG model, there are a total of 19 layers. The VGG19 model comes to a close with three more FC levels, bringing the total number of layers to 19. Each of these layers has 4096, 4096, and 1000 neurons respectively.

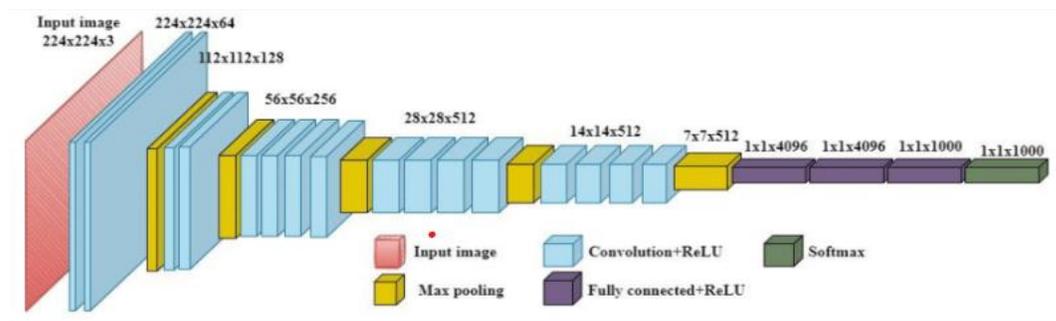

Figure 4: VGG19 Model Architecture

In addition to that, there are five Maxpool layers and a Softmax layer that are provided. The activation of ReLU is a feature that is present in layers that are convolutional in nature [28].

### 3.4.3 DenseNet 121

Dense connections between layers are prioritized in the DenseNet121 convolutional neural network design. Researchers at Facebook AI Research created the 121-layer neural network known as DenseNet121, which has convolutional, pooling, and fully connected layers. Unlike traditional CNN designs, DenseNet121 connects each layer to each other in a feed-forward manner, allowing for a direct flow of data and gradients across the network.

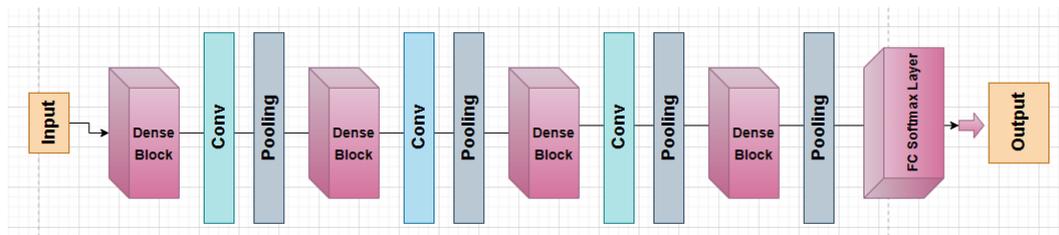

Figure 5: DenseNet 121 Model Architecture

This thick connection improves gradient-based optimization and boosts model performance by encouraging feature reuse, reducing the number of parameters, and improving gradient flow. Due to its exceptional performance in picture categorization and object identification tasks, DenseNet121 is a preferred choice for many computer vision applications [29].

### 3.4.4 ResNet50

A deep convolutional neural network architecture called ResNet50 has made major strides in computer vision. It has 50 layers and residual blocks that help the network get around the difficulties associated with training very deep neural networks.

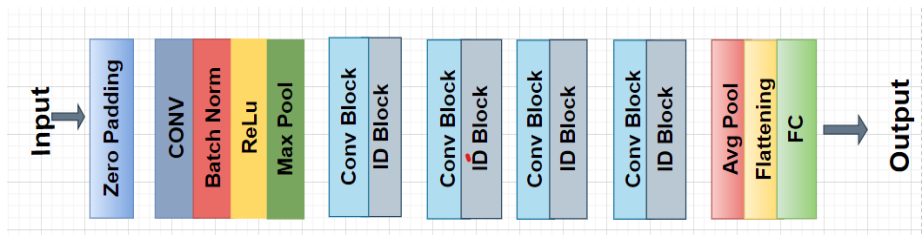

Figure 6: ResNet50 Model Architecture

ResNet50 has demonstrated outstanding performance in a variety of tasks, including image segmentation, object identification, and classification. Its skip connections provide improved gradient flow and avoid network performance from declining as network depth increases by allowing information to flow straight from early levels to later layers. ResNet50 has grown to be a popular option in several cutting-edge deep learning applications because of its potent feature extraction capabilities [30].

### 3.4.5 YOLO V4

YOLOv4 is a cutting-edge object identification algorithm that is famous for the extraordinary precision and speed with which it operates. It is an acronym for "You Only Look Once" and is utilized frequently in many computer vision jobs. YOLOv4 is equipped with sophisticated features such as a CSPDarknet53 backbone and various scale predictions, which give it the ability to identify objects of varying sizes and scales in an effective manner.

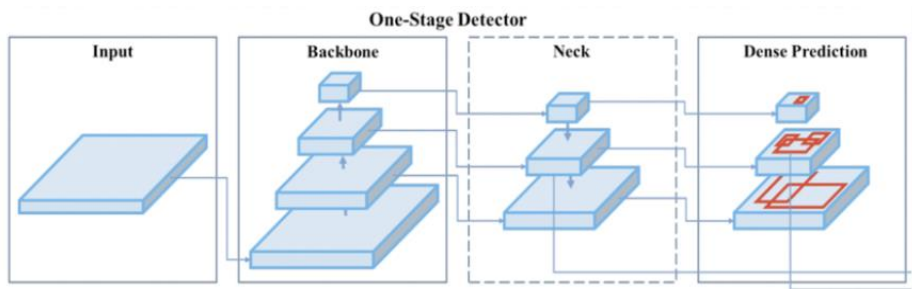

Figure 7: YOLO V4 Model Architecture

In addition to this, it implements the Mish activation function and presents a number of optimizations, such as spatial pyramid pooling (SPP) and panoptic feature pyramid networks (PFPN). In general, YOLOv4 has earned a stellar reputation for the reliable performance it delivers in real-time object detection applications [31].

# CHAPTER 4

# RESULT ANALYSIS

## 4.1 Introduction

The result analysis section presents a comprehensive evaluation of the proposed model's performance. It includes the evaluation metrics used to measure the performance of the model, such as accuracy, precision, recall, and F1 score. The impact on the model's performance is also discussed, along with the confusion matrix to assess the model's ability to correctly classify each class. Additionally, the performance of the proposed model is compared with five other conventional transfer learning models to determine its superiority. Overall, this section provides a detailed analysis of the proposed model's performance and its strengths and weaknesses compared to other models.

## 4.2 Evaluation Metrics

The accuracy, losses, and performance parameters of each of the five transfer learning models are examined in the outcome analysis. Based on its performance analysis, which takes accuracy, recall, F1 score, and other pertinent parameters into account, the optimal model is chosen. The best model for soil recognition may be found thanks to this thorough evaluation, which also offers insights into how well it performs overall. Precision, recall, F1-score, accuracy (ACC), sensitivity, and specificity determined the optimal model. Models have confusion matrices. Thus, TPs, TNs, FPs, and FNs are identified. FPR, FNR, FDR, MAE, and RMSE were computed for model statistical analysis.

$$\text{Accuracy} = (TP + TN)/(TP + TN + FP + FN) \quad (1)$$

$$\text{Recall} = TP/(TP + FN) \quad (2)$$

$$\text{Specificity} = TN/(TN + FP) \quad (3)$$

$$\text{Precision} = TP/(TP + FP) \quad (4)$$

$$\text{F1 score} = (2 \times \text{Precision} \times \text{Recall})/(\text{Precision} + \text{Recall}) \quad (5)$$

$$FPR = FP/(FP + TN) \quad (6)$$

## 4.3 Result of Transfer Learning Model

Table 3 illustrates the five transfer learning models' training, test, validation, and loss. The table shows that the ResNet50 model is the most accurate.

**Table 3.** Result of Transfer learning model

| Model | Train_Acc. | Train_Loss | Val_Acc. | Val_loss | Test_Acc. | Test_loss |
|---|---|---|---|---|---|---|
| VGG19 | 96.63 | 0.21 | 95.93 | 0.21 | 95.22 | 0.25 |
| VGG16 | 96.21 | 0.20 | 96.95 | 0.12 | 96.12 | 0.20 |
| DenseNet 121 | 97.41 | 0.19 | 97.23 | 0.28 | 97.21 | 0.31 |
| **ResNet50** | **99.98** | **0.23** | **99.54** | **0.32** | **99.54** | **0.37** |
| YOLO V4 | 91.23 | 0.39 | 91.21 | 0.392 | 91.21 | 0.94 |

Confusion metrics, commonly referred to as confusion matrices, are an important technique for assessing how well categorization algorithms work. For a particular dataset, they offer a thorough overview of the projected and actual class labels. The matrix is set up as a square grid, with the genuine classes represented by rows and the anticipated classes by columns. Each matrix column displays the number or percentage of cases that belong to a certain combination of true and anticipated classes. Confusion metrics include a number of indicators, including accuracy, precision, recall, and F1 score, that help analyze the model's strengths and shortcomings, notably in terms of misclassifications and the trade-off between various sorts of mistakes. These metrics aid in evaluating and improving classification models, allowing data scientists and practitioners to decide on the effectiveness of the model and any room for improvement. Important visualizations that shed light on the effectiveness and development of machine learning models are loss and accuracy curves. The value of the loss function, such as cross-entropy or mean squared error, is plotted against the quantity of training epochs or iterations on the y-axis to create the loss curve. As it quantifies the difference between the expected and actual values, the objective is to minimize the loss. The accuracy curve, on the other hand, shows how the model's accuracy changes with the number of epochs or iterations on the x-axis. Better performance is indicated by higher accuracy values. Data scientists may determine if the model is converging, overfitting, or underfitting the data by examining these graphs. By choosing the right number of training epochs, seeing possible problems, and making wise model improvement decisions, these visualizations aid in model optimization.

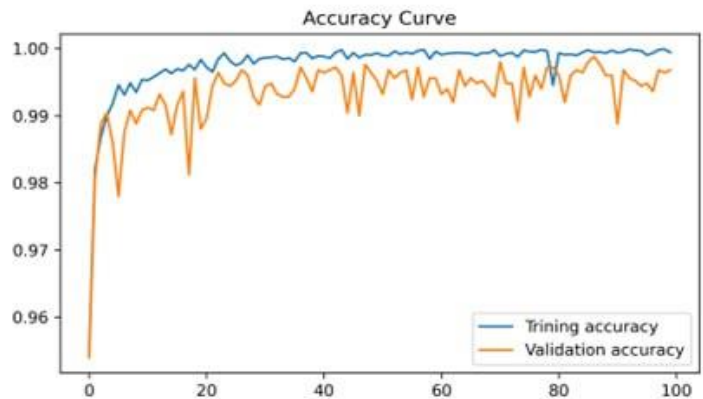

Figure 8: Accuracy curve over 100 epochs

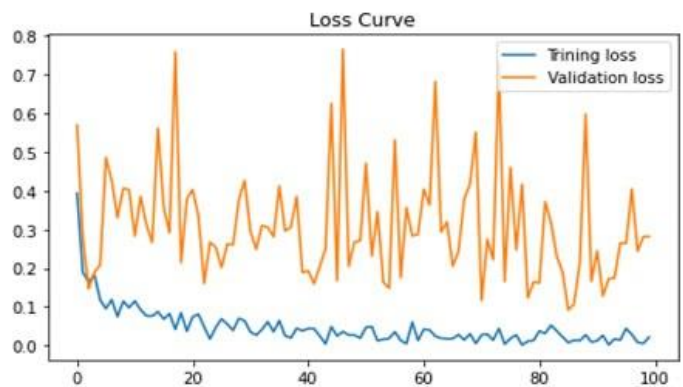

Figure 9: Loss curve over 100 epochs

## 4.4 Performance analysis and statistical analysis

Showing the FDR, FPR, FNR, KC, MCC, MAE, and RMSE values of the best model (ResNet50) in table 4.

Table 4. Performance Analysis and Statistical Analysis

| Accuracy | FPR (%) | FNR (%) | FDR (%) | KC (%) | MCC (%) | MAE | RMSE |
|---|---|---|---|---|---|---|---|
| 99.54 | 0.12 | 0.46 | 0.46 | 98.95 | 88.62 | 0.58 | 8.85 |

The performance measures of the best model, ResNet50, are presented in Table 4. These metrics include the False Discovery Rate (FDR), the False Positive Rate (FPR), the False Negative Rate (FNR), the Matthews Correlation Coefficient (MCC), the Kappa Coefficient (KC), the Mean Absolute Error (MAE), and the Root Mean Square Error (RMSE). These metrics give insights into the accuracy, precision, and overall performance of ResNet50 in the identification of brain tumors, allowing for a full evaluation and comparison of its efficacy against the performance of other models or benchmarks.

In tumor detection, we have used five transfer learning models. These are VGG16, VGG19, DenseNet121, ResNet50, and YOLO V4. Among them, ResNet50 achieved the highest accuracy of 99.54%. Table 5 shows the Performance Analysis of Precision, Recall, and F1-Score for RestNet50.

Table 5. Performance Analysis of Precision, Recall and F1-Score

| Class | Precision (%) | Recall (%) | F1-Score (%) | Accuracy (%) |
|---|---|---|---|---|
| ResNet50 | 99.54 | 99.54 | 99.54 | 99.54 |

The accuracy of a model's positive predictions is measured by precision, also known as Positive Predictive Value (PPV). A high accuracy number shows that the model does an excellent job of avoiding false positive predictions, which means that it seldom incorrectly labels negative events as positive. Recall, often referred to as Sensitivity or True Positive Rate (TPR), gauges how well a model can pick out instances of positivity among the entire number of real occurrences of positivity. When there is an imbalance between positive and negative cases in the dataset, the F1-Score is helpful. The F1-value has a range of 0 to 1, with 1 representing the ideal value. It performs best when accuracy and recall are evenly distributed.

# CHAPTER 5
# IMPACT ON SOCIETY, ENVIRONMENT AND SUSTAINABILITY

## 5.1 Introduction

In this part of the study, we look into the influence that the research has had on society as a whole, as well as the environment and the capacity to support it. We investigate how the application of deep learning techniques in the diagnosis of brain tumors might provide major advantages to society, including improvements in healthcare outcomes, greater diagnostic accuracy, and prompt treatments. In addition, we explore the possible environmental consequences of our findings, such as a reduced dependence on invasive procedures and unneeded imaging tests, which will lead to more sustainable healthcare practices and a lower environmental imprint in the area of medical imaging.

## 5.2 Impact on Society

The use of deep learning methods to the diagnosis of brain tumors will have a significant and far-reaching effect on society. It is possible to dramatically increase both the accuracy and the efficiency of diagnosing brain tumors by making use of more complex algorithms, such as ResNet50. This will ultimately lead to improved medical outcomes for patients. Interventions can be performed at the appropriate moment after early and correct discovery of brain tumors, which increases the likelihood of successful treatment and may save lives. Additionally, the automation and simplification of the detection process that may be accomplished with deep learning models can lighten the load on healthcare workers, enabling them to devote more time and resources to the care of their patients. The application of deep learning to the identification of brain tumors has a beneficial influence on society as a whole since it improves medical diagnostics, makes it possible to develop individualized treatment plans, and eventually leads to an improvement in the wellbeing and quality of life of people whose lives have been damaged by brain tumors.

## 5.3 Impact on the environment

The use of methods including deep learning in the diagnosis of brain tumors has a further beneficial effect on the surrounding natural environment. It is possible to eliminate the necessity for invasive treatments such as biopsies by making use of more sophisticated algorithms such as ResNet50. This not only makes patients feel less discomfort and lessens the hazards connected with invasive operations, but it also results in less trash being produced by medical facilities. In addition, techniques that are based on deep learning make it possible to do imaging that is both more precise and more focused. This cuts down on the number of scans that aren't necessary and the related consumption of resources like energy, materials, and chemicals. Deep learning leads to a more sustainable and environmentally aware approach to the identification of brain tumors and medical imaging as a whole by improving the diagnostic process and encouraging more efficient use of resources. This is accomplished through the promotion of efficient resource use.

## 5.4 Ethical Aspects

When it comes to the diagnosis of brain tumors, the application of deep learning algorithms raises a number of important ethical questions. Protecting the privacy of patients and their data should be considered one of the most important ethical considerations. It is necessary to maintain tight confidentiality, acquire informed permission, and follow to data protection standards in order to preserve patient information while using deep learning models because these models depend on vast volumes of patient data. Another ethical issue to consider is the bias and fairness of algorithms. Deep learning models have the potential to inherit biases from the data on which they are trained, which might result in diagnostic and treatment discrepancies for some segments of the population. It is important to make an effort to identify and reduce any bias that may exist in the models in order to guarantee that they are fair, objective, and relevant to a wide range of patient groups. Additionally essential to ethical deliberation are factors such as openness and explicability. It might be difficult to comprehend how deep learning models, such as ResNet50, come to the conclusions that they do because these models are sometimes referred to as "black boxes."

It is crucial to ensure that the models are interpretable and explainable in order to build trust, which will promote clinical adoption and enable clinicians to comprehend and dispute the model's findings. Consenting after being fully informed and participating in decision-making together are both ethical values that should be upheld. Patients should be given the chance to participate in the decision-making process and should be sufficiently informed about the use of deep learning models in their diagnosis. Patients should also be given appropriate information on the use of deep learning models in their diagnosis. Clinicians have a responsibility to give patients with thorough explanations and assist them in making educated decisions on their treatment options based on the results produced by deep learning models. It is essential to keep monitoring, validating, and improving deep learning models in order to make certain that they continue to be effective while also being safe. Audits and evaluations should be carried out on a regular basis in order to evaluate the performance of the models, as well as their correctness and any potential biases. In order to preserve the health and safety of patients and the public's faith in the technology, any flaws or restrictions should be pinpointed and immediately corrected. Ethical issues also extend to the responsibilities of healthcare professionals and researchers to utilize deep learning models as tools to complement clinical decision-making rather than to replace human knowledge. This is important because of the potential for unethical behavior if this responsibility is not met. It is important to remember that physicians bear the ultimate responsibility for patient care and should base their judgments not only on their own knowledge but also on the results produced by the deep learning model. Another component of ethics is making sure that everyone has the same opportunities. In order to prevent further aggravating existing healthcare inequities, it is imperative that deep learning models be made available for use in a wide variety of healthcare settings, including those with limited access to resources. It is important that steps be taken to close the digital gap and ensure that everyone has equal access to the advantages offered by deep learning technology. Ethical issues also include the deployment of deep learning models in a responsible and transparent manner. In order to cultivate scientific rigor, reproducibility, and accountability, adequate documentation, the sharing of methodology, and peer review are all necessary components. Open communication and teamwork between physicians, researchers, and policymakers, as well as between patients and researchers, can assist resolve ethical concerns and encourage responsible innovation. In conclusion, the fast developing area of deep learning calls for a continuous debate on ethics as well as ethical advice in order to successfully manage future ethical difficulties. In order to address the one-of-a-kind ethical issues that are raised by deep learning models in the context of brain tumor detection, ethical frameworks, guidelines, and regulatory frameworks need to be regularly updated. The ethical use of deep learning in brain tumor diagnosis can maximize its advantages while simultaneously limiting potential hazards and assuring patient-centered treatment if certain ethical norms are upheld and active obstacles connected with the usage are actively addressed.

## 5.5 Sustainability Plan

A sustainability strategy for the application of deep learning in the diagnosis of brain tumors should incorporate a number of essential components. To begin, it should make the responsible management of resources a top priority by improving computing algorithms and infrastructure in order to reduce the amount of energy used and waste produced. In addition, the strategy need to encourage the creation of and acceptance of open-source and transparent frameworks that make it possible for researchers to collaborate, share their expertise, and reproduce their findings. It is essential to maintain deep learning models' usefulness and impact over the long run by providing for their ongoing maintenance and support. In addition, the plan need to advocate for the ethical acquisition and usage of data, putting an emphasis on privacy, security, and consent as the highest priorities. Finally, the promotion of multidisciplinary collaborations and partnerships between academia, industry, healthcare providers, and policymakers may promote the sustainable integration of deep learning in brain tumor detection. This will ensure that the technology will continue to bring long-term advantages to society as well as environmental considerations.

# CHAPTER 6
# CONCLUSION

## 6.1 Introduction

In this part of the article, we will show the findings and conclusions of our research project on the application of transfer learning models to the identification of brain tumor on brain MRI dataset. In addition, we emphasize prospective topics for further investigation and acknowledge the constraints that were encountered throughout the course of the research process. This part gives a detailed summary of the study's outcomes by summarizing the findings, addressing the limits, and recommending future directions. It also paves the way for additional developments in the field of brain tumor detection based on MRI images research by setting the stage for these developments.

## 6.2 Conclusion

In conclusion, the use of deep learning strategies, in particular ResNet50, has shown tremendous potential in the field of detecting brain tumors. When comparing tumor instances with non-tumor cases, the use of ResNet50, which has deep layers and robust feature extraction capabilities, has demonstrated exceptional accuracy and efficiency in making the distinction between the two types of cases. Researchers have been able to construct reliable and automated methods for accurate brain tumor identification by training ResNet50 on big datasets of brain MRI images. This has allowed the researchers to develop the systems more quickly. The use of deep learning strategies, such as ResNet50, in the analysis of medical images offers the potential to improve the speed, accuracy, and objectivity of diagnosing brain tumors. This is one of the many areas in which deep learning has shown promise. Deep learning-based approaches have the potential to alter the area of brain tumor identification and contribute to improved patient outcomes if additional improvements are made in the technology behind these methods and research into them is maintained.

## 6.3 Limitation and Future Work

In spite of the substantial progress that has been made in detecting brain tumors through the use of deep learning algorithms, there are still certain limits and areas that need further research. One of the limitations is the requirement for extensive annotated datasets that are both broad and varied in order to guarantee the generalizability of the model across a variety of patient groups and cancer types. Another obstacle to overcome is the difficulty of interpreting the results of deep learning models, which are sometimes referred to as "black boxes." If this problem were solved by the creation of explainable deep learning algorithms, the level of confidence and acceptance that these models would get in therapeutic contexts would increase. Integration of multi-modal data, such as combining MRI with other imaging modalities or clinical data, might further increase the accuracy and reliability of brain tumor identification. Other imaging modalities include PET, CT, and SPECT. Continued research efforts should focus on resolving these constraints, enhancing deep learning algorithms, and undertaking prospective clinical trials to test the performance and effectiveness of deep learning-based brain tumor detection systems in real-world settings. These are all important areas of attention.

# APPENDIX: Research Reflections

Before I began, I did not know anything about artificial intelligence or deep learning. My supervisor is a great person who is kind and honest. She was very helpful and gave important advice right from the start. I gained a lot of knowledge while doing the research, like how to make a better set of data and how to get useful information from data that is not organized. How to use and apply algorithms and other related techniques.I applied five transfer learning models in this research project. These are VGG16, VGG19, DenseNet 121, ResNet50, and YOLO V4 where ResNet50 provide the best accuracy model which is 99.54%. Python is the most widely used in my research project. I use Jupiter Notebook, Google Colab and Spyder for coding or implementation. In the end,I studied AI, deep learning, and computer vision. This has motivated me to continue learning more about these subjects in the future.